\documentclass[apj]{emulateapj}
\usepackage{lineno}
\usepackage[ngerman,english]{babel}
\newif\ifAMStwofonts
\AMStwofontstrue

\def\arcsec{\hbox{$^{\prime\prime}$}}

\def\bz{{b(z)}}
\def\nz{{N(z)}}

\def\clth{{C^{\rm th}_\ell}}

\def\clcorr{{C^{\rm corrected}_\ell}}

\def\al{{a_{\ell m}}}

\def\gsim{~\rlap{$>$}{\lower 1.0ex\hbox{$\sim$}}}

\def\simpropto{\lower.2ex\hbox{$\; \buildrel \propto \over \sim \;$}}
\def\ltsim{\lower.5ex\hbox{$\; \buildrel < \over \sim \;$}}
\def\gtsim{\lower.5ex\hbox{$\; \buildrel > \over \sim \;$}}
\def\ltsim{\lower.5ex\hbox{$\; \buildrel < \over \sim \;$}}
\def\gtsim{\lower.5ex\hbox{$\; \buildrel > \over \sim \;$}}

\def\dd{\,{\rm d}}

\def\vb{{v_{_{\rm b}}}}

\def\dd{{\rm d}}

\def\ln{{\rm ln}}

\def\pmb#1{\setbox0=\hbox{#1}%
\kern-.025em\copy0\kern-\wd0
\kern.05em\copy0\kern-\wd0
\kern-.025em\raise.0433em\box0}

\def\vr{\pmb{r}}

\def\hvn{\hat {\vr}}

\def\vk{\pmb{k}}

\def\simlt{\lower.5ex\hbox{$\; \buildrel < \over \sim \;$}}
\def\simgt{\lower.5ex\hbox{$\; \buildrel > \over \sim \;$}}

\newcommand{\beq}{\begin{equation}}

\newcommand{\eeq}{\end{equation}}
\def\beqa{\begin{eqnarray}}
\def\eeqa{\end{eqnarray}}
\def\fixit#1{}

\def\dd{{\rm d}}

\def\cN{{\cal N}}

\usepackage{float}
\usepackage{amsmath}
\usepackage{bm}
\usepackage{ulem}
\usepackage{color}
\usepackage{graphicx}
\usepackage{hyperref}
\usepackage{enumerate}
\usepackage{soul,xcolor}
\usepackage{cancel}
\usepackage{physics}
\usepackage{multirow}
\setstcolor{red}

\begin{document}
\title{ Galaxy power spectrum and biasing results from the LOFAR Two-metre Sky Survey (first data release)}
\author{Prabhakar Tiwari}
\email{ptiwari@nao.cas.cn}
\affil{National Astronomy Observatories,
Chinese Academy of Science, Beijing, 100101, P.R.China}

\author{Ruiyang Zhao}
\affil{National Astronomy Observatories,
Chinese Academy of Science, Beijing, 100101, P.R.China}
\affil{University of Chinese Academy of Sciences, Beijing 100049, P.R.China}

\author{Jinglan Zheng}
\affil{National Astronomy Observatories,
Chinese Academy of Science, Beijing, 100101, P.R.China}
\affil{University of Chinese Academy of Sciences, Beijing 100049, P.R.China}
\affil{Institute of Cosmology and Gravitation, University of Portsmouth, Dennis Sciama Building, Portsmouth PO1 3FX, United Kingdom}

\author{Gong-Bo Zhao}
\affil{National Astronomy Observatories,
Chinese Academy of Science, Beijing, 100101, P.R.China}
\email{gbzhao@nao.cas.cn}
\affil{School of Astronomy and Space Science, University of Chinese Academy of Sciences, Beijing 100049, P.R.China}

\author{David Bacon}
\affil{Institute of Cosmology and Gravitation, University of Portsmouth, Dennis Sciama Building, Portsmouth PO1 3FX, United Kingdom}

\author{Dominik J. Schwarz} 
\affil{Fakult{\"a}t f{\"u}r Physik, Universit{\"a}t Bielefeld, Postfach 100131, 33501 Bielefeld, Germany}
%\author{} \email{}\affil{}
%\date{\today}
\begin{abstract}
The LOFAR Two-metre Sky Survey (LoTSS) is an ongoing survey aiming to observe the entire Northern sky, providing an excellent opportunity to study the distribution and evolution of the large-scale structure of the Universe. The source catalogue from the public LoTSS first data release (DR1) covers 1\% of the sky, and shows correlated noise or fluctuations of the flux density calibration on few degree scales. We explore the LoTSS DR1 to understand the survey systematics and data quality of this  first data release. We produce catalog mocks to estimate uncertainties, and measure the angular clustering statistics of LoTSS galaxies, which fit the $\Lambda$CDM cosmology reasonably well. We employ a Markov chain Monte Carlo (MCMC) based Bayesian analysis to recover the best galaxy biasing scheme and multi-component source fraction for LoTSS DR1 above $1$ mJy assuming different possible redshift templates. After masking some noisy and uneven patches and with suitable flux density cuts, the LOFAR survey appears qualified for large-scale cosmological studies. The upcoming data releases from LOFAR are expected to be deeper and wider, and will therefore provide improved cosmological measurements. 
\end{abstract}
\maketitle

\section{Introduction}
 Our present understanding of the origin and evolution of the Universe is based on the Lambda cold dark matter ($\Lambda$CDM) cosmology. In this model, the matter density of the Universe is dominated by cold dark matter, whose gravitational evolution results in a population of virialized  dark matter halos of different masses \citep{Press:1974,Scherrer:1991,Navarro:1997,Ma:2000,Seljak:2000,Scoccimarro:2001,Cooray:2002}. The formation of galaxies occurs inside these dark matter halos, and the host halo mass and its evolution correlates with the evolution and type \citep{Girelli:2020} of galaxy residing inside. In general,  radio-loud galaxies' active galactic nuclei (AGNs) are found to reside in more massive halos than optical AGNs \citep{Mandelbaum:2008,Wilman:2008}. That being so, the optical and radio observations sample quite a different set of galaxies (halos). Most galaxies which are bright at optical wavelengths are undetectable at radio wavelengths, and strong radio sources are often optically faint or invisible. The radio surveys sample the higher end of the halo mass range as compared with optical observations, and thus complement existing and upcoming visible/IR galaxy surveys \citep{Kauffmann:2003,Best:2005,Mandelbaum:2009,Best:2014,Adi:2015nb,Krumpe:2018,Hale:2018,Alonso:2020jcy,Lan:2021,Wolf:2021}. 

The international Low-Frequency Array (LOFAR; \citealt{Haarlem:2013}) is a new-generation radio interferometer constructed in northern Netherlands and across Europe, offering increase in radio survey speed with unparalleled sensitivity and angular resolution in the low-frequency radio regime \citep{Rottgering:2003,Haarlem:2005,Rottgering:2005,Falcke:2007}. The LOFAR Two-metre Sky Survey (LoTSS) is an ongoing deep $120-168$ MHz imaging survey being carried out using LOFAR high-band antenna (HBA) observations across the whole northern hemisphere \citep{Shimwell:2017,Shimwell:2019}. The LoTSS aims to explore cosmic large-scale structure, galaxies, clusters of galaxies, and the formation and evolution of massive black holes. The LoTSS survey will observe millions of radio AGNs,  along with a significant number of star-forming galaxies (SFGs) out to redshift $z\sim 6$, allowing detailed studies of the physics and evolution of AGNs and SFGs. The LoTSS is producing high-fidelity images at a central frequency of $144$ MHz with a resolution of $6\arcsec$ and with declination-dependent sensitivity, typically around $100~\mu {\rm Jy}$/beam. This is a factor of $10$ more sensitive than previous high-resolution sky surveys, e.g., VLA's FIRST. LoTSS will ultimately detect over $10$ million radio sources with a significant fraction of star-forming galaxies. A large fraction of LoTSS sources will have optical identifications and photometric redshifts will be available \citep{Williams:2019,Duncan:2019}. Furthermore, the  WHT Enhanced Area Velocity Explorer (WEAVE) multi-object spectrograph on the William Herschel Telescope will observe optical ($370-970$ nm) spectra of millions of LOFAR radio sources and provide precise redshift information \citep{Smith:2016}. 

The LoTSS survey, homogeneously covering the whole northern sky complete down to the sub mJy limit  will overcome statistical limitations due to shot noise. The large galaxy number density and large sky coverage  will substantially reduce cosmic variance in cosmological analysis. 
The radio galaxies, tracing the background dark matter, will constrain the shape of power spectrum, i.e., the early universe physics, dark matter, baryon, and neutrino densities, the inflation power spectrum and the degree of non-Gaussianity in density fluctuations. The upcoming LoTSS catalogs, covering a large sky area, will help us to explore further regarding large-scale anomalies \citep{Oliveira-Costa:2004,Ralston:2004,Schwarz:2004,Tiwari:2019l123} and the current puzzling dipole signal observed with radio catalogs \citep{Blake:2002, Singal:2011,Gibelyou:2012,Rubart:2013,Tiwari:2014ni,Tiwari:2015np, Tiwari:2016adi, Colin:2017,Siewert:2020CRD}. Furthermore, LoTSS will significantly improve on present low-frequency radio catalogs, e.g., TIFR GMRT Sky Survey (TGSS; \citealt{Intema:2016tgss}) and GaLactic and Extragalactic All-sky MWA (GLEAM; \citealt{Hurley:2017gleam}), 
and analyses based on these surveys \citep{Tiwari:2016tgss,Rana:2019, Dolfi:2019,Tiwari:2019TGSS,Choudhuri:2020}. Unfortunately, the link between the galaxy and total matter power spectra depends on some unknowns from astrophysics such as the galaxy bias factor, which depends on galaxy type  and is quite different for radio AGNs  and star-forming galaxies. The LoTSS population is a mixture of AGNs and star-forming galaxies, and therefore understanding galaxy bias, relative number densities and luminosity evolution is non-trivial. The purpose of this work is to present a detailed cosmological analysis of LoTSS galaxies and study the effect of survey footprint, shot-noise and other systematics. We have produced galaxy mocks for the survey and have customized and calibrated the data pipeline for galaxy clustering statistics recovery. 

We introduce the catalog for LoTSS DR1, its completeness and data mask in Section \ref{sc:catalog} and \ref{ssc:mask}. In Section \ref{sc:covariance}, we present mock generation details and covariance matrix estimation. We briefly present the theoretical formulation of the galaxy angular power spectrum and its connection to the dark matter power spectrum in Section \ref{sc:galaxyPk}. In Sections \ref{sc:cl_measure} and \ref{sc:2pcf}, we present our measured angular power spectrum and two-point correlation statistics, respectively. We present an estimate for the bias for the LoTSS population in Section \ref{sc:bias_nz} and summarize our results in Section \ref{sc:summary}. We conclude with discussion in Section \ref{sc:discussion}. 

\section{LoTSS DR1 catalog}
\label{sc:catalog}
The LOFAR Two-metre Sky Survey (LoTSS; \cite{Shimwell:2017}) is ongoing and plans to scan the entire northern sky at 120-168 MHz. \cite{Shimwell:2019} have prepared the first full-quality public data release (LoTSS-DR1) catalog from  $63$
LoTSS  data sets (2\% of the total survey) in the region of the HETDEX Spring Field that were observed between 2014 May 23 and 2015 October 15. The LoTSS DR1 has been prepared using a fully automated direction-dependent calibration and imaging pipeline discussed in \cite{Shimwell:2017}. The catalog covers $424$ square degrees and contains a total of $325,694$ sources with peak flux density at least five times the local rms noise, thus a source density of about $770$ sources per square degree. The resolution of the survey images is $6\arcsec$ and the positional accuracy of the sources in the catalog is within $0.2\arcsec$. The median sensitivity is $71~\mu {\rm Jy}$/beam  at $144$ MHz. \cite{Williams:2019} remove artefacts, correct wrong groupings of Gaussian components and prepare a value-added catalog; the catalog then contains $318,520$ sources of which $231,716$ have optical/near-IR identifications in Pan-STARRS\footnote{\href{https://www.ifa.hawaii.edu/research/Pan-STARRS.shtml}{https://www.ifa.hawaii.edu/research/Pan-STARRS.shtml}}/WISE\footnote{\href{https://www.nasa.gov/mission_pages/WISE/main/index.html}{https://www.nasa.gov/mission\_pages/WISE/main/index.html}
}. Not all of these have photo-$z$ detection, as Pan-STARRS is not as complete as WISE and we only have photo-$z$s for about 50\% of all radio sources.
\begin{figure*}
    \centering
    \includegraphics[width=1.0\textwidth]{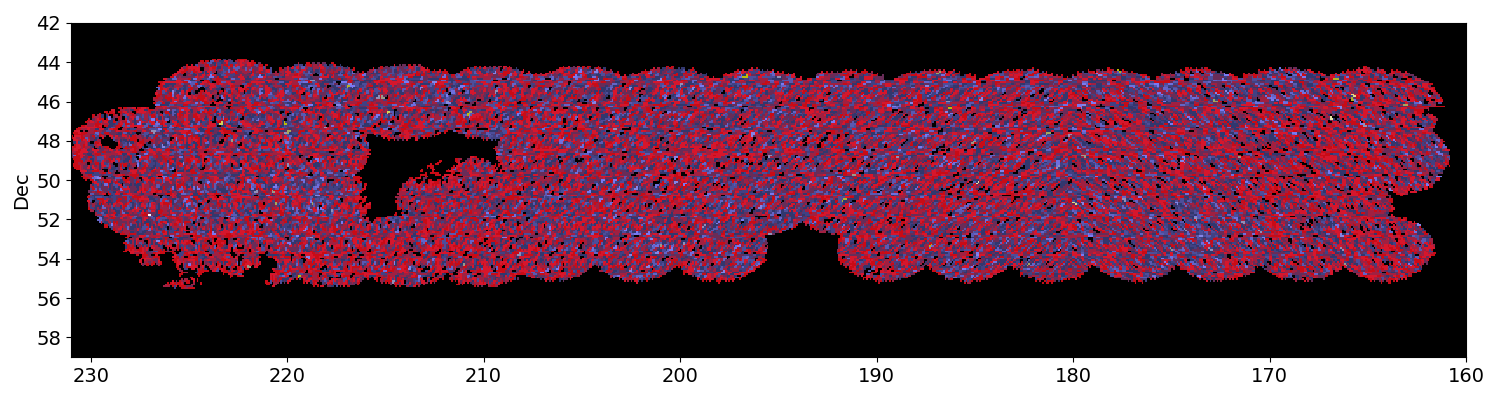}\\
     \includegraphics[width=1.0\textwidth]{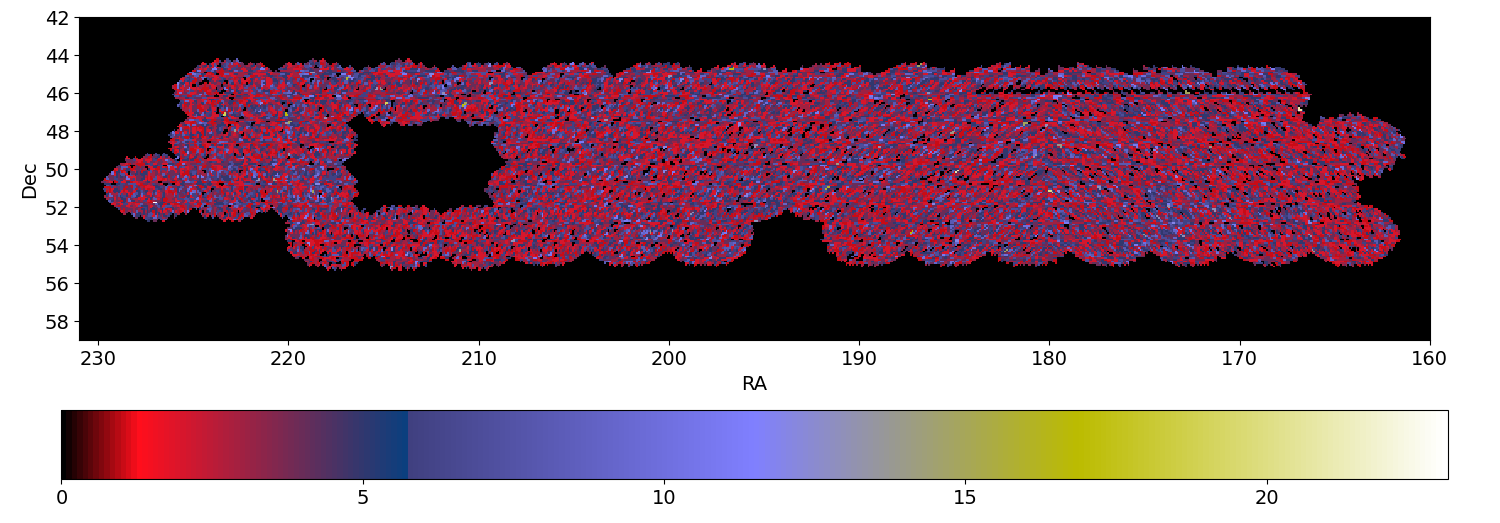}\\
    \caption{The distribution of radio galaxies in the LoTSS-DR1 catalog with integrated flux density between 1 to 10,000 mJy. Plotted are the number counts in pixels in Cartesian projection at HEALPix\footnote{\href{http://healpix.sourceforge.net }{http://healpix.sourceforge.net }} resolution Nside = 512 before (top) and after(bottom) applying the mask. After applying the mask and 1-10,000 mJy flux cut, there are 107926 sources remaining.}
    \label{fig:LoTSS}
\end{figure*}

\subsection{Completeness and data mask}
\label{ssc:mask}
\begin{figure*}
    \centering
    \includegraphics[width=1.1\textwidth]{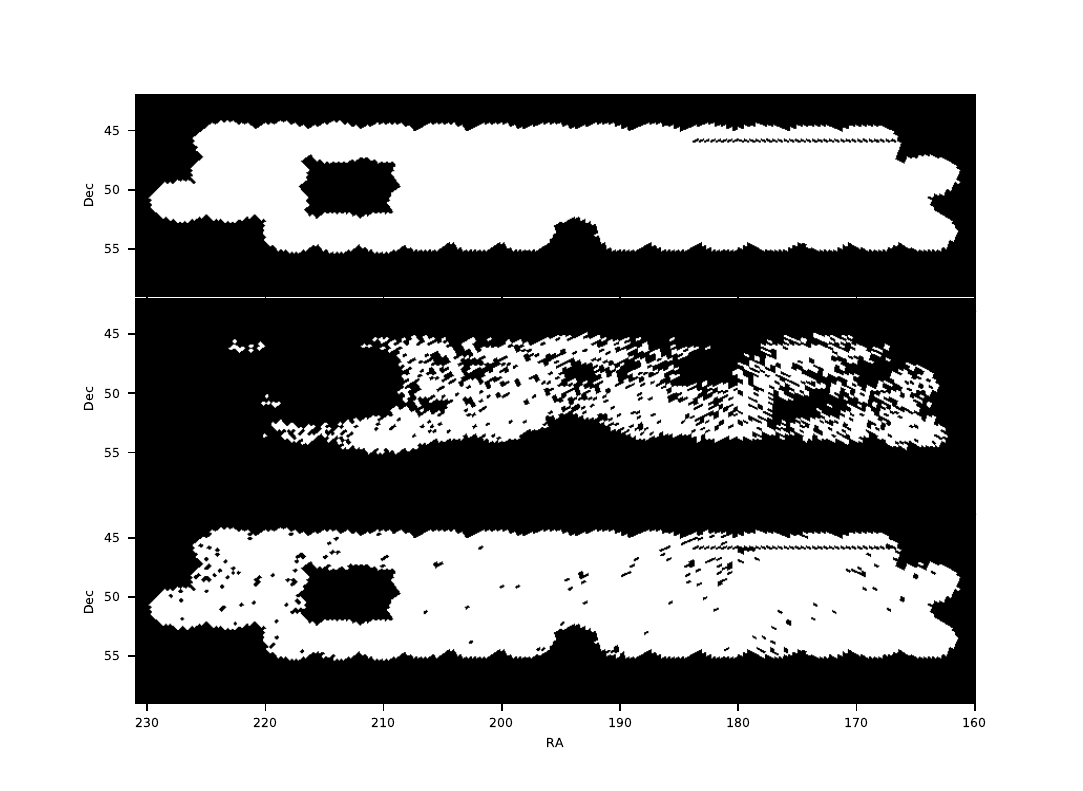}
    \caption{LoTSS DR1 mask-z, z1 and z2. The mask-z (top) rejects 5 under-sampled pointings and includes all 53 good pointings of the LOFAR HETDEX field. Also it rejects regions of equal declination, where information from Pan-STARRS is missing. The mask has been prepared considering the survey geometry, consistency of source counts, and sampling uniformity \citep{Siewert:2019}. In the middle and bottom panel we show mask-z1 and mask-z2 which are obtained by further masking cells with a local noise above the median noise and two times the median noise, respectively. }
    \label{fig:mask}
\end{figure*}

\cite{Shimwell:2019} estimate the completeness of LoTSS-DR1 and report the catalog to be $65\%$ complete at $0.23$ mJy, $90\%$ at $0.45$ mJy, and $95\%$ at $0.58$ mJy. The catalog is more than 99\% complete at $1$ mJy. The LoTSS DR1 catalogue was generated by combining $58$ individual LOFAR pointings on the sky, and due to poor ionospheric conditions and due to the presence of bright sources, the imaging pipeline occasionally produces sub-standard images. As a result some pointings show inhomogeneous point source distribution, for example in particular 5 pointings show significant incompleteness \citep{Siewert:2019}. After properly defining the survey region and masking $5$ incomplete pointings, a reasonable survey mask (mask-z) for reliable cosmology is shown in figure \ref{fig:mask}. Mask-z is an upgrade to mask-d \citep{Siewert:2019}; it further rejects regions where information from Pan-STARRS is missing. In addition to mask-z we further consider masking cells with a local noise above the median noise (mask-z1) and two times the median noise (mask-z2)\footnote{We thank Thilo Siewert for preparing and making mask-z,z1,z2 available to us.}. We impose the completeness flux cut (i.e., 1 mJy) and also remove the ultra bright sources with flux equal to or  greater than $10$Jy; the catalog thus contains $121,730$ sources. After employing mask-z shown in figure \ref{fig:mask}, $107,926$ sources remain.  

\section{Mock catalogs and Covariance matrix estimation}
\label{sc:covariance}
Mock catalogues are essential for assessing the analysis pipeline, data systematics and errors in the cosmological analysis of large galaxy surveys. In order to estimate the uncertainty in the cosmological signal recovered from LoTSS data, we generate $1000$ LoTSS mocks of the large-scale structure by employing the log-normal (and Gaussian) density field simulator code FLASK\footnote{\href{http://www.astro.iag.usp.br/~flask/}{http://www.astro.iag.usp.br/~flask/}} \citep{Xavier:2016}. 
To emulate the LoTSS DR1 catalog, we generate multiple log-normal density fields tomographically in $35$ redshift slices ($0$ to redshift $3.5$), each with a width $\Delta z=0.1$. All statistical properties (i.e., auto and cross-correlations) are determined by the input angular power spectrum. In addition, effects including redshift-space distortions and lensing are also included in the simulation via the input spectra provided to the FLASK pipeline. 

The input theoretical angular power spectrum, $C_\ell$, in different redshift bins are generated using {\tt CAMB} \citep{Challinor:2011} which includes the effects on the observed number counts of redshift-space distortions, non-linear power spectrum corrections and lensing. The redshift distribution profile $N(z)$ is provided  along with $C_\ell$ to FLASK to generate mock number count maps. 
Computation of theoretical $C_\ell$ requires a galaxy bias, $b(z)$. From our MCMC fits we iteratively  provide best fitted bias and generate mock LoTSS DR1 catalogs. We assume a $\Lambda$CDM model and use \cite{Planck_results:2018} cosmological parameters as the fiducial cosmology throughout this work, in particular we set base cosmology parameters from Planck 2018 baseline, i.e. TT,TE,EE+lowE+lensing $\omega_b h^2 =0.02237$,    $\omega_c h^2  =0.1200$,  $100\theta_{MC}=1.04092$, $\tau=0.0544$,   $\ln (10^{10} A_s)=3.044$ and $n_s=0.9649$.

\begin{figure}
    \centering
    \includegraphics[width=0.5\textwidth]{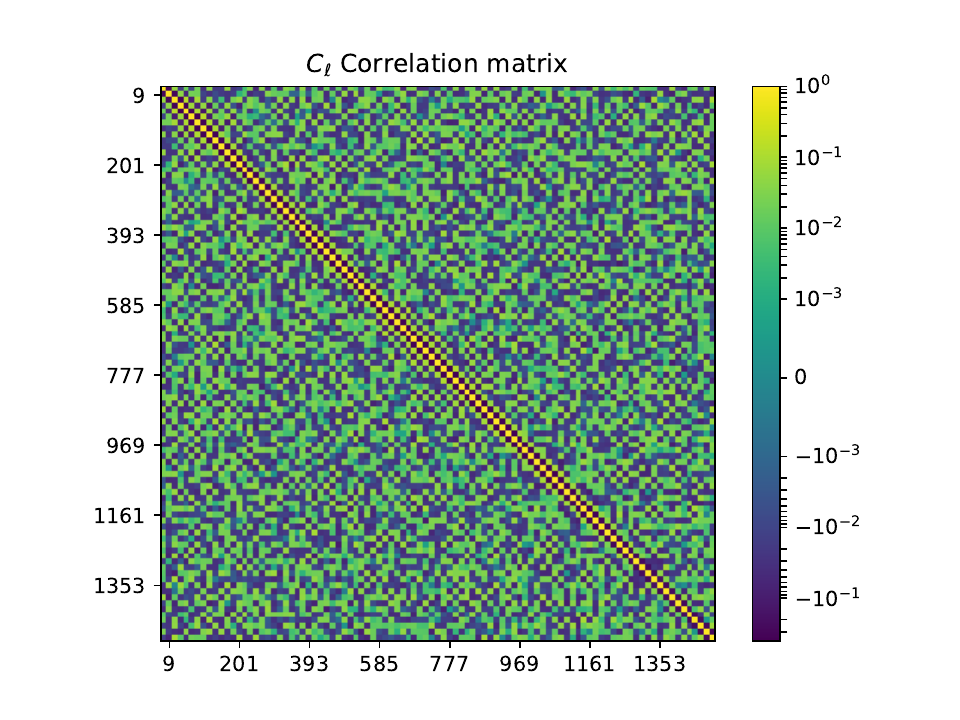}
    \caption{Angular power spectrum correlation matrix constructed from 1000 mocks. The galaxy mocks are generated using the FLASK log-normal density simulator. Each galaxy mock follows the same sky coverage and contains approximately the same number of sources as LoTSS DR1 after mask and flux cuts.}
    \label{fig:covmat}
\end{figure}

We apply the LoTSS DR1 survey mask to precisely account for survey geometry, then from the mock catalogs we calculate the angular power spectrum up to $l\approx 1500$. Although LoTSS DR1 is a relatively high number density radio catalog, it is hardly over 1\% of the sky, and so, given the small sky coverage, $C_\ell$ is noisy.  Therefore we choose $\Delta {\ell}=16$ and  recover $C_\ell$ in bands by collecting 16 multipoles per bin. Finally, with mock $C_\ell$, we compute a covariance matrix to determine the uncertainty in LoTSS DR1 recovered galaxy power. A plot of the correlation matrix of the binned angular power spectrum is shown in figure \ref{fig:covmat}.

\section{ Galaxy Angular power spectrum }
\label{sc:galaxyPk}
The galaxies are biased tracers of the underlying dark matter density and thus the underlying cosmological model. The theoretical relationship between the total matter density perturbations and galaxy density are easily  written in terms of galaxy clustering measures. The statistical measure of clustering of galaxies can be conventionally expressed in terms of the angular power spectrum, $C_\ell$. Following the $\Lambda$CDM scenario, the theoretical formulation of $C_\ell$ is briefly as follows.
Assume a uniform galaxy survey catalog with an area $\mathcal{A}$ and total number of galaxies $\cN$. The mean number density, $\bar \cN$, per steradian is thus simply $\cN/\mathcal{A}$. Subsequently, $\cN(\hvn) =\bar \cN(1+\Delta (\hvn))$, is  the projected number density per steradian in  the direction $\hvn$. 
Here $\Delta (\hvn)$ is the projected number density contrast and is theoretically connected to the underlying matter density contrast, $\delta_m(\vr,z(r))$. Note that $\vr$ stands for comoving distance $r$ in direction $\hat{r}$ and $z(r)$ is the redshift corresponding to comoving distance $r$.
The galaxy density contrast, $\delta_g(\vr,z(r))$, in the direction $\hvn$ and at redshift $z$, in terms of matter density contrast $\delta_m(\vr,z(r))$, 
\beq
\delta_g(\vr,z(r)) =\delta_m(\vr,z=0) D(z) b(z),
\eeq
where $b(z)$ is galaxy bias and $D(z)$ is the linear growth factor. Following these we can formulate the theoretical $\Delta (\hvn)$,
\begin{eqnarray}
\label{eq:delta_th}
\Delta (\hvn) &=& \int _{0}^{\infty} \delta_g(\vr, z(r)) p(r) dr  \nonumber\\
              &=&  \int _{0}^{\infty} \delta_m(\vr,z=0)  D(z) b(z) p(r) dr ,
\end{eqnarray}
where $p(r) \dd r$, the radial distribution function, is the probability of observing a galaxy between comoving distance $r$ and $(r+ \dd r)$. Note its connection to redshift distribution profile $N(z)$, $p(r) \propto N(z) \dd z/\dd r$. The $\Delta (\hvn)$ may also have some tiny additional contributions from lensing, redshift distortions, physical distance fluctuations and from variation of radio source luminosities and spectral indices \citep{Chen:2015}. However, these effects are expected to be limited to a few percent on the largest scales \citep{Dolfi:2019}. Next, we expand $\Delta (\hvn)$ in terms of spherical harmonics to obtain $C_{\ell}$,
\beq
\label{eq:alm}
\Delta (\hvn) = \sum_{\ell m}\al Y_{\ell m}(\hvn).
\eeq

We use the orthonormal property of spherical harmonics and write $\al$ as
\begin{eqnarray}
\label{eq:alm_gal}
\al&=&\int d \Omega \Delta(\hvn)  Y_{\ell m}^*(\hvn)\\
\nonumber &=& \int d\Omega Y_{\ell m}^*(\hvn) \int_{0}^{\infty} dr \delta_m(\vr,z=0)  D(z) b(z) p(r)\;.
\end{eqnarray}

We can Fourier transform the matter density field $\delta_m(\vr,z=0)$ in terms of the $\vk$-space density field $\delta_{\vk}$, 
\begin{equation}
\label{eq:deltak}
\delta_m(\vr,z=0) = \frac{1}{(2\pi)^3} \int d^3 k \delta_{\vk} e^{i \vk .\vr},
\end{equation}
and substitute 
\begin{equation}
    e^{i \vk .\vr} = 4 \pi \sum_{\ell,m} i^{\ell} j_{_\ell}(kr) Y_{\ell m} (\hvn) Y^{*}_{\ell m}(\hat{\vk}), 
\end{equation}
where $j_{_\ell}$ is the spherical Bessel function of the first kind for integer $\ell$. Next, we can write $\al$ as
\begin{equation}
\label{eq:alm_th2}
\al=\frac{{i}^\ell}{2\pi^2}\int dr D(z) b(z) p(r) \int d^3 k \delta_{\vk}j_\ell(kr) Y^*_{\ell m}(\hat {\vk}) \; . 
\end{equation}

Subsequently, we find  the expression for the angular power spectrum, $C_\ell$, as 
\begin{eqnarray}
\label{eq:clth}
C_\ell&=&<|\al |^2> \nonumber\\
\nonumber &=& \frac{2}{\pi }\int dk k^2 P(k) \left\vert \int_{0}^{\infty} d r D(z) b(z) p(r)  j_\ell(kr)\right\vert^2 \\
          &=&   \frac{2}{\pi }\int dk k^2 P(k) W^2(k) \;,
\end{eqnarray}
where $P(k)$ is the matter power spectrum, and $W(k)=\int_{0}^{\infty} d r  D(z) b(z) p(r)  j_\ell(kr)$ is the $k$-space window function. The galaxies  are discrete point sources and thus the measured angular power spectrum of a galaxy catalog incorporates the Poissonian shot-noise contribution $\sigma_0$, equal to $\frac{1}{\bar \cN}$. In addition a radio catalog in general contains multiple entries for some sources causing a constant offset to $C_\ell$. This offset can be approximated in terms of shot-noise $\sigma_0$. Therefore, the corrected power spectrum, $\clcorr$, corresponding to the theoretical $C_\ell$ given in Equation (\ref{eq:clth}) is $C^{\rm measured}_\ell - \sigma_0 \times \alpha$, where $\alpha$ is 
1 (shot-noise) + an offset factor due to the multi-component correction. We have discussed this further in section \ref{sc:cl_measure}. The uncertainty in  $C_\ell$ determination due to cosmic variance, sky coverage and shot-noise is
\begin{equation}
\label{eq:dcl}
\Delta C_\ell = \left( \frac{2}{(2\ell+1)f_{\rm sky}}\right)^{1/2} C^{\rm measured}_\ell,
\end{equation}
where $f_{\rm sky}$ is the fraction of sky covered by  survey. 

\section{\texorpdfstring{$C_\ell$}{cl} measurements}
\label{sc:cl_measure}
\begin{figure}
    \centering
    \includegraphics[width=0.5\textwidth]{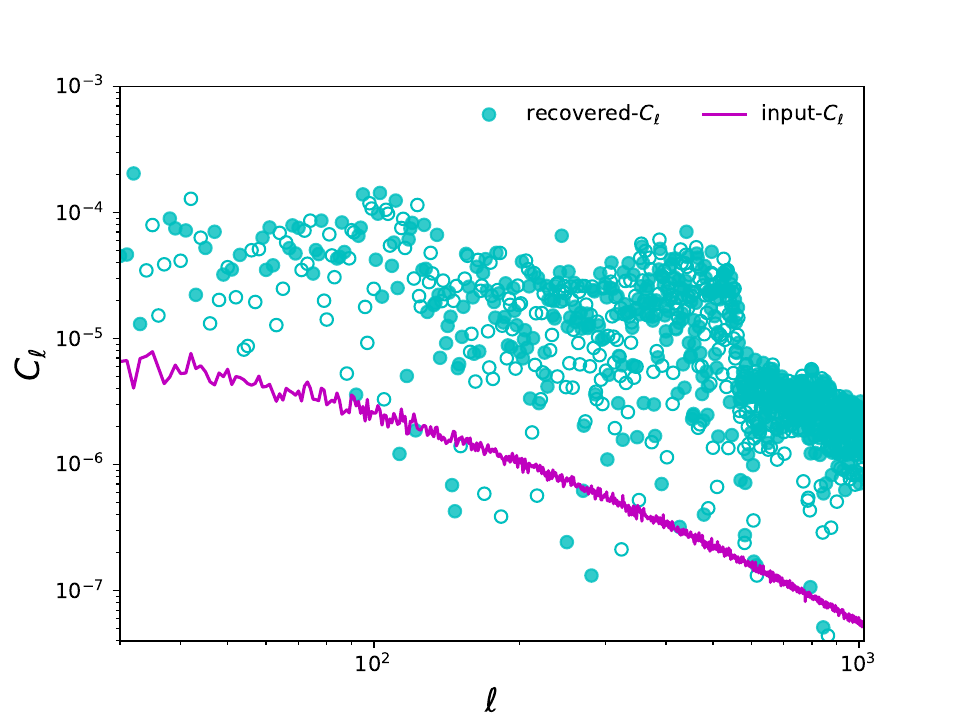}\\
    \includegraphics[width=0.5\textwidth]{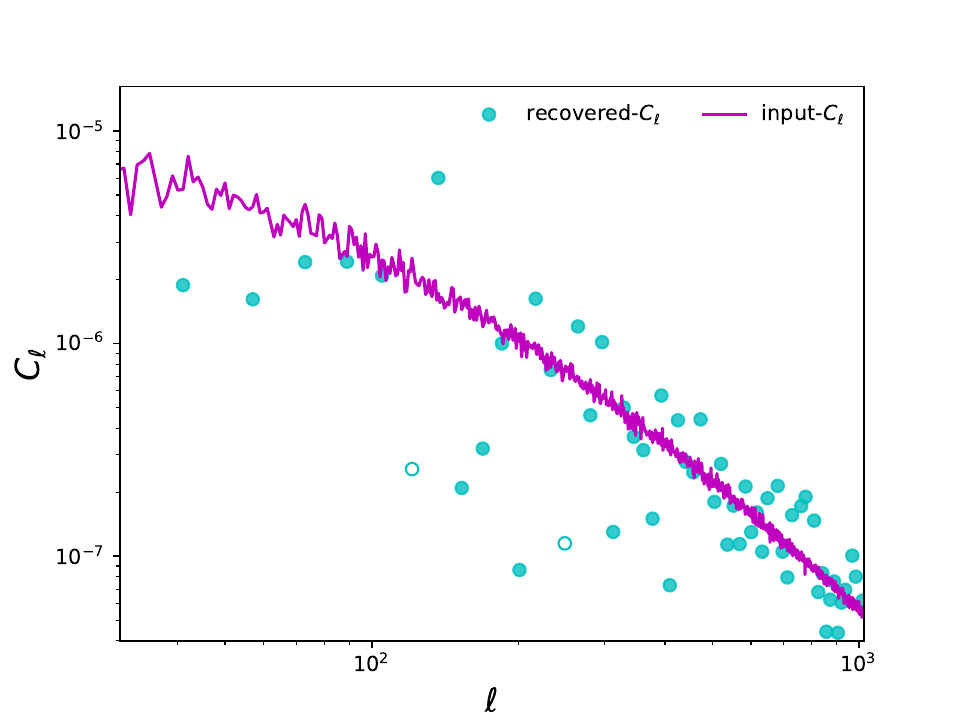}\\
    \includegraphics[width=0.5\textwidth]{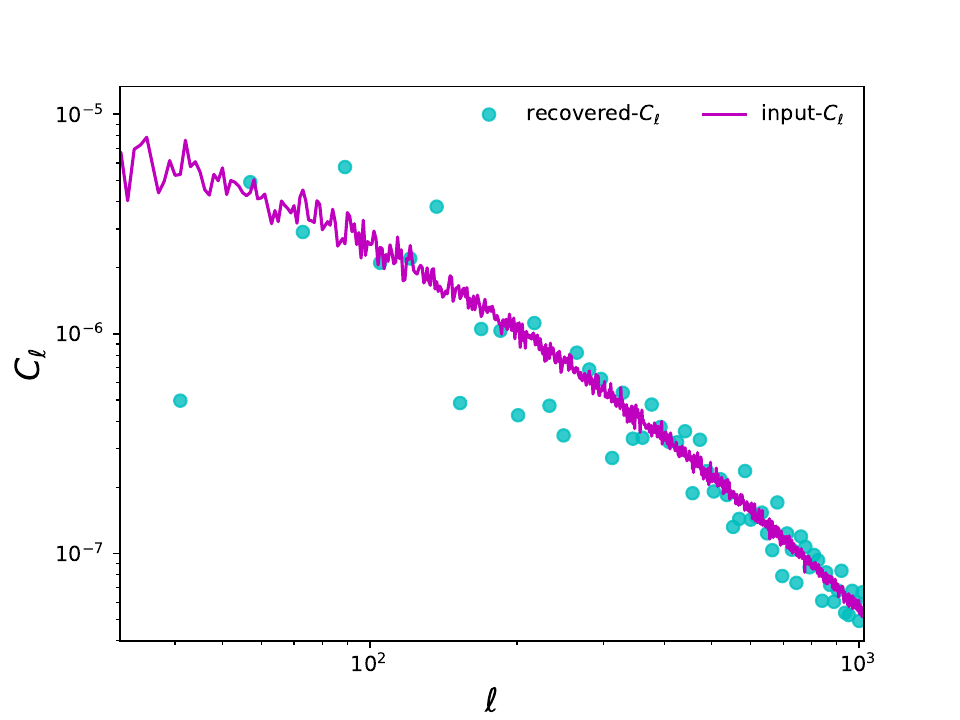}\\
    \caption{Angular power spectrum recovery performance with LoTSS DR1 mask ($0.1\%$ of the sky). Empty circles show a negative value. The recovered power spectrum is significantly fluctuating as a function of multipole (top figure). Next we apodized the mask on the scale of one degree and plot $C_\ell$ in bands by collecting 16 multipoles per bin; we show the results for this in the middle panel. The best recovery above $\ell >100$ is obtained if we do not apodize the mask and collect 16 multipoles per bin (bottom panel).}
    \label{fig:cl_recovery}
\end{figure}

We make use of a pseudo-$C_\ell$ recovery algorithm  by \cite{Alonso:2019}  to achieve an efficient and reasonably accurate estimate of the angular power spectrum of the LoTSS DR1 catalog. The algorithm python module is publicly available as NaMaster\footnote{\href{https://namaster.readthedocs.io/en/latest/index.html}{https://namaster.readthedocs.io/en/latest/index.html}}, and the mathematical background of the estimator, its features, and software implementation are described in \cite{Alonso:2019}. We test the performance of the pseudo-$C_\ell$  algorithm for the LoTSS DR1 mask  using a test map, which resembles the LoTSS DR1 density fluctuations as follows. We obtain a reasonable recovery of $C_\ell$ by using an un-apodized mask shown in figure \ref{fig:mask}. To emulate the LoTSS DR1 catalog, we first generate  $\al$ using  LoTSS DR1 model $C_\ell$ \footnote{$C_\ell$ is the expected variance of the $\al$ at $\ell$.}; subsequently we obtain a density contrast map from $\al$ by making use of spherical harmonics (equation \ref{eq:alm}). Next, we apply the LoTSS DR1 mask to this test map and recover $C_\ell$ (pseudo-$C_\ell$s). The results thus obtained with different configurations and settings are shown in figure \ref{fig:cl_recovery}, and above $\ell>100$ a reasonable recovery is demonstrated.

\begin{figure}
    \centering
    \includegraphics[width=0.5\textwidth]{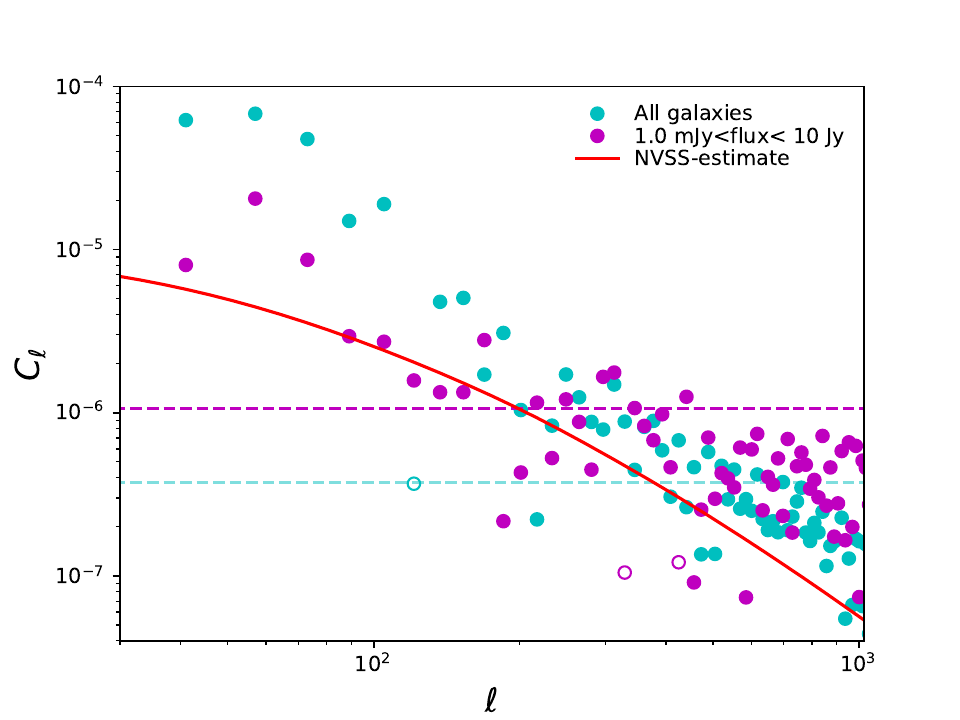}
    \caption{LoTSS DR1 angular power spectrum recovery. We obtain an approximate theoretical power spectrum `NVSS-estimate' for LoTSS DR1 catalog following equation \ref{eq:clth} and assuming NVSS galaxy bias and redshift distribution \citep{Adi:2015nb}. The power spectrum obtained without a flux cut is significantly high at large scales (low $\ell$). The dashed lines show the shot-noise. The LoTSS DR1 catalog is only expected to be complete above 1 mJy.}
    \label{fig:dataCl}
\end{figure}

\begin{figure}
    \centering
    \includegraphics[width=0.5\textwidth]{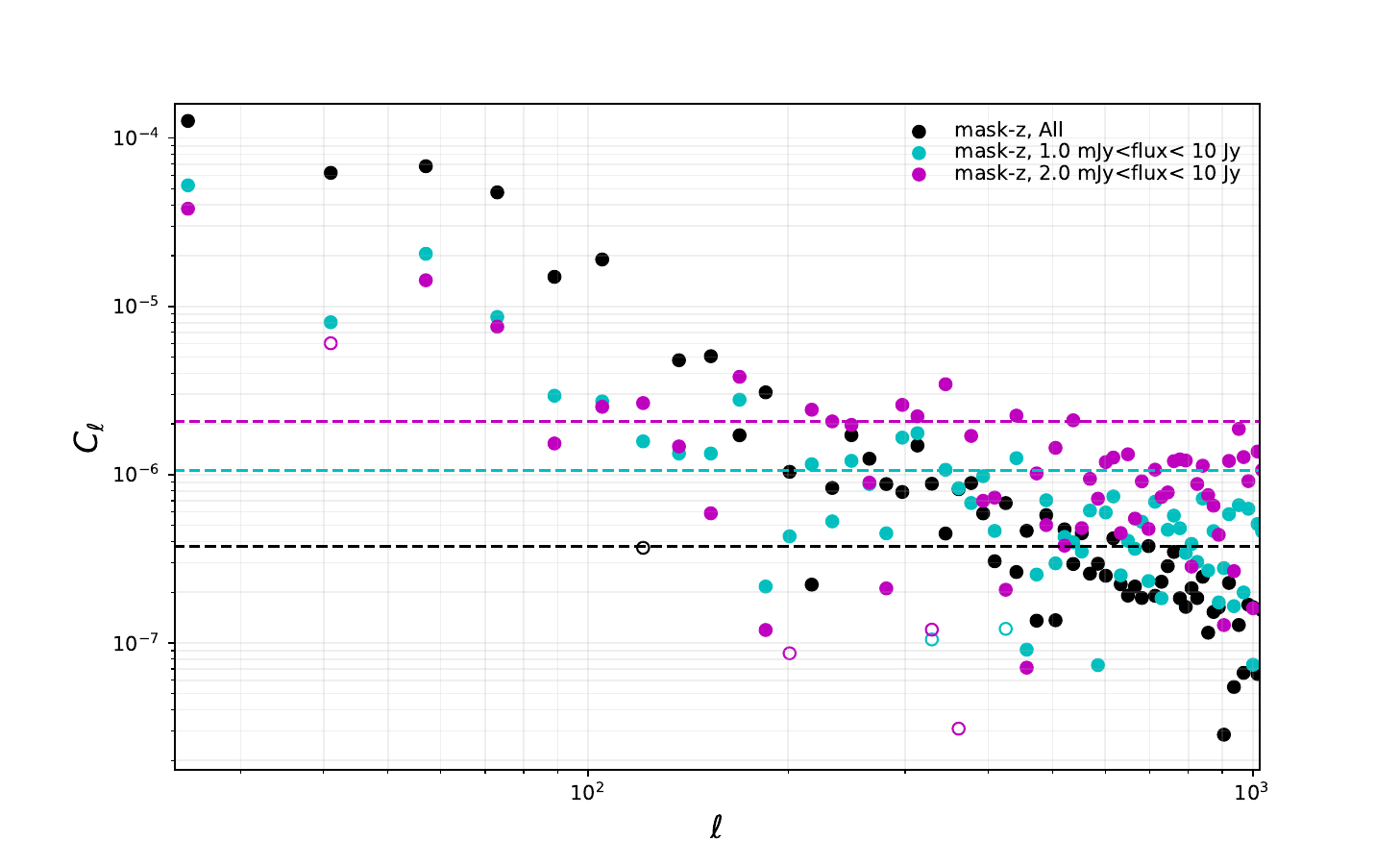}
    \caption{LoTSS DR1 angular power spectrum recovery with different flux cuts.The empty circles show the negative value. The dashed lines show the shot-noise. }
    \label{fig:Cl_fluxs}
\end{figure}

\begin{figure}
    \centering
    \includegraphics[width=0.5\textwidth]{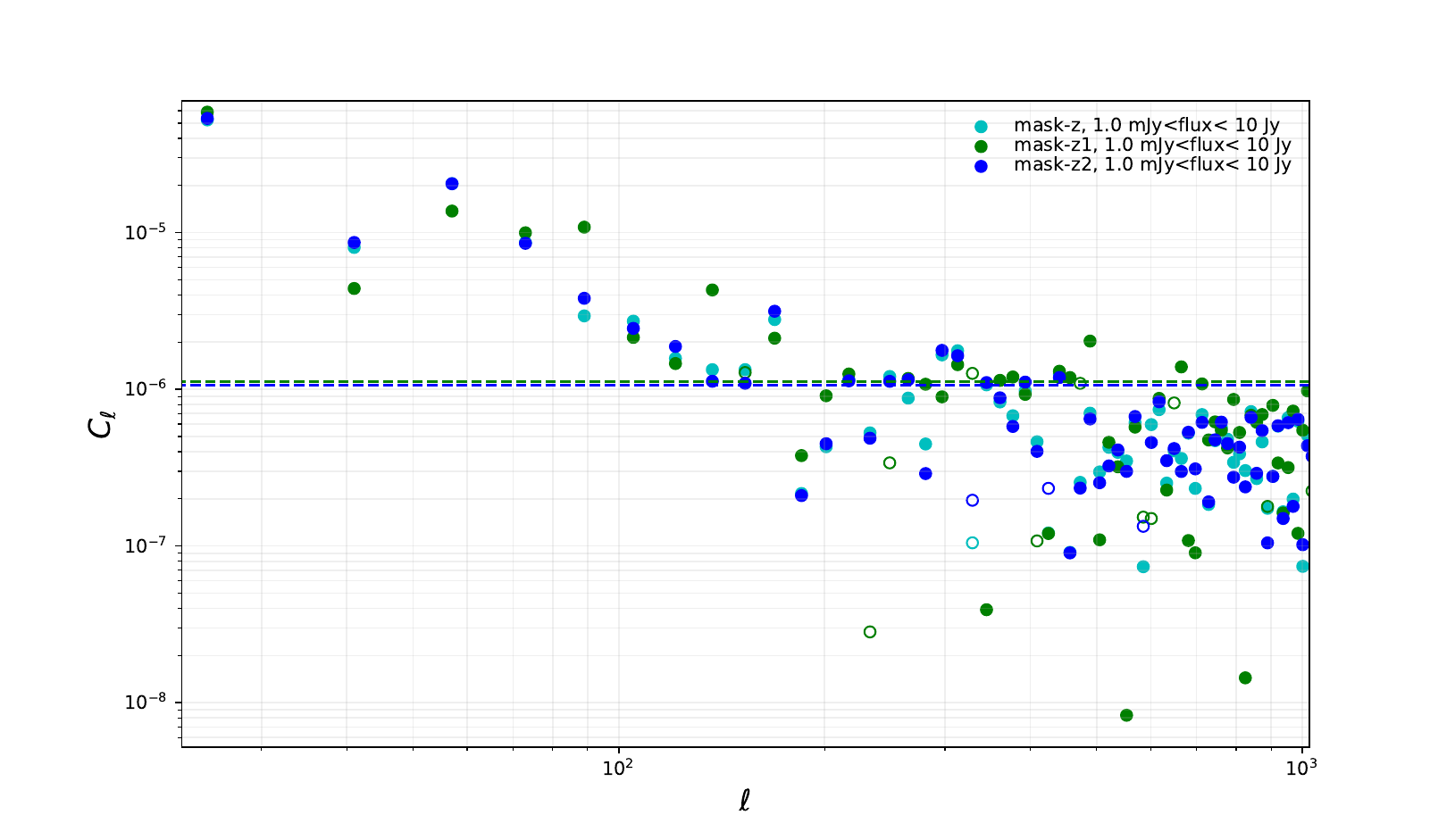}
    \caption{LoTSS DR1 angular power spectrum recovery with mask-z,z1, and z2. These masks are described in section \ref{ssc:mask} and shown in figure \ref{fig:mask}. Other details same as figure \ref{fig:Cl_fluxs}. }
    \label{fig:Cl_masks}
\end{figure}

With the above demonstrated settings, we run the pseudo-$C_\ell$ recovery algorithm and  obtain the LoTSS DR1 catalog angular power spectrum. The angular spectrum thus obtained is shown in figure \ref{fig:dataCl}. The recovered power spectrum for galaxies with integrated  radio flux  above survey completeness, i.e. $S>1$ mJy, approximately agrees with theoretical results obtained following $\Lambda$CDM and considering NVSS \citep{Condon:1998} radio galaxy biasing  and radial distribution \citep{Adi:2015nb}. Without considering survey completeness (i.e. the flux cut) the recovered power spectrum is significantly high at low $\ell$, i.e. at large scales, presumably attributing the effect of  survey incompleteness. At low fluxes, when the survey is not complete, we may have large scale flux systematics and uneven source counts and thus may observe excess large scale clustering. Considering a slightly higher flux cut, i.e. $S>2$mJy, we obtain similar results as for the $S>1$ mJy flux cut $C_\ell$ but with slightly more fluctuations.  The results are shown in figure \ref{fig:Cl_fluxs}. We next test the robustness of $C_\ell$ recovery with different masks discussed in section \ref{ssc:mask}. With mask-z, z1 and z2 we recover almost the same $C_\ell$, although the fluctuations for mask-z1 are slightly higher due to low sky coverage (i.e., high cosmic variance). The results are shown in figure \ref{fig:Cl_masks}. 

Depending  on the flux density threshold considered, the LoTSS DR1 catalog contains a significant number of multi-component sources \citep{Siewert:2019}.  The multi-component sources cause a constant offset to $C_{\ell}$. \cite{Siewert:2019} estimate the clustering parameter $n_c$ for LoTSS DR1, which is defined to be the ratio of the variance of counts in cells over the mean of the counts in cells. In addition they notice that the counts-in-cells statistic follows a compound Poisson distribution, assuming that the parameter $n_c=1+\gamma$, where $\gamma$ is the mean number of components for a source. Given this, for any source in the catalog the probability of having one or two components can be given as $\gamma \exp(-\gamma)$, or $\frac{\gamma^2}{2!} \exp(-\gamma)$, respectively. This implies that the ratio between the number of two-component and one-component sources equals $\gamma/2$, and thus the fraction of radio sources split into double sources i.e. $e$ as defined in \cite{Blake:2004}, equals $\frac{\gamma/2}{1+\gamma/2}$. We further note from \cite{Siewert:2019} that $\gamma=0.44$ for 1 mJy flux density threshold, and estimate the multi-component offset in terms of shot-noise $\sigma_0$ to equal $\gamma \sigma_0/(1+\gamma) \approx 0.3 \sigma_0$ for $C_\ell$ following \cite{Blake:2004}. The estimate of course makes several assumptions as discussed above and is not guaranteed to be exact. We thus consider the multi-component offset to $C_\ell$s as an independent parameter in our fitting procedure. Indeed, we recover a multi-component offset close to $0.3 \sigma_0$ as expected.

\section{2-point correlation statistics}
\label{sc:2pcf}
\begin{figure}
    \centering
    \includegraphics[width=0.5\textwidth]{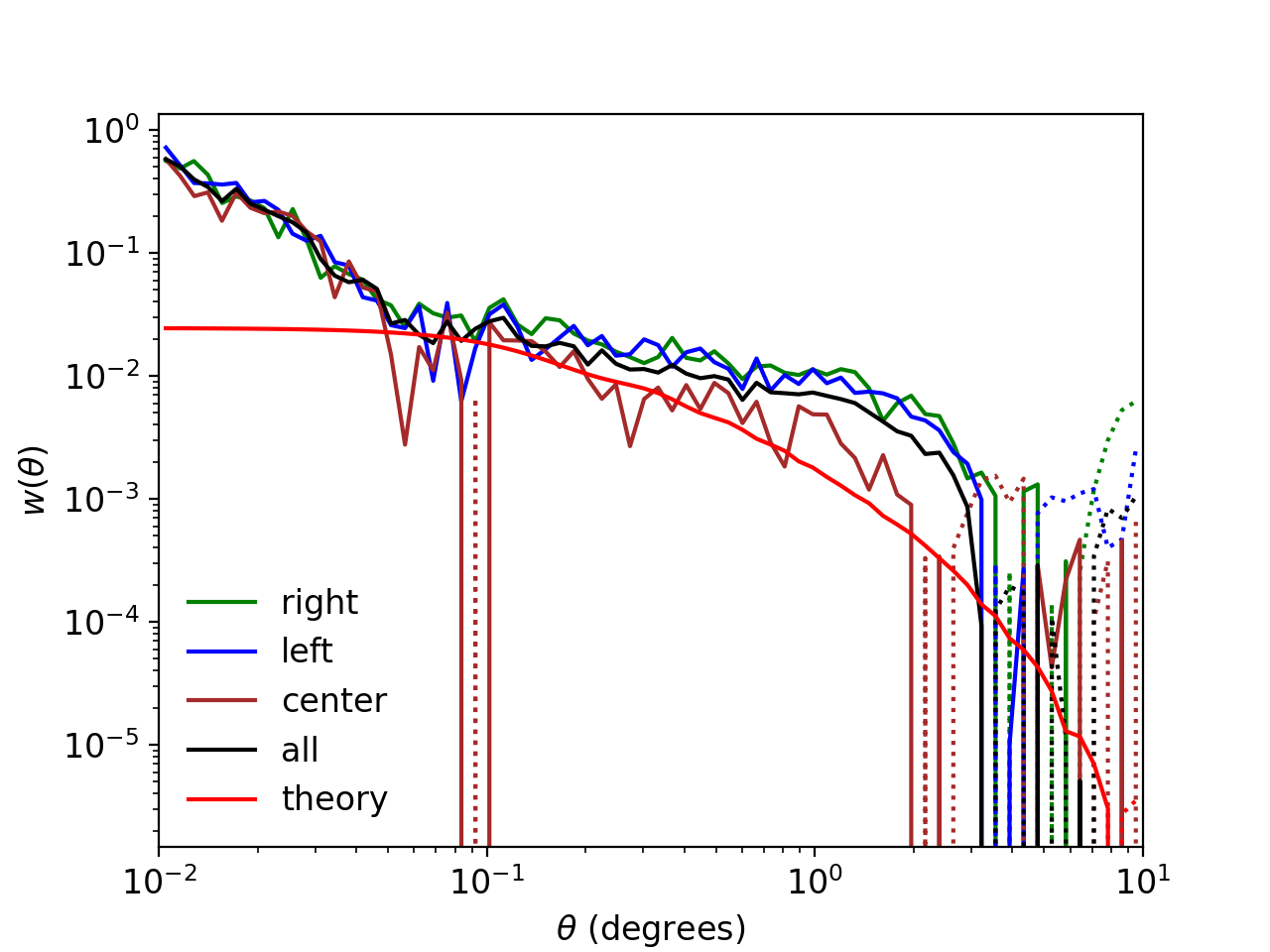}
    \caption{The 2-point correlation function of LoTSS DR1 catalogues in different patches of the sky. The green curve denotes the `right' patch angular correlation function, the blue for `left' area and orange for `center' area. The whole area angular correlation function of LoTSS DR1 is shown by the black line, while the red curve denotes the theoretical prediction following $\Lambda$CDM with NVSS bias and redshift distribution. All dotted lines represent negative values.}
    \label{fig:2pcf}
\end{figure}

For comparison with theory and earlier results in \cite{Siewert:2019}, we calculate the 2-point correlation function (2PCF) for the LoTSS DR1 catalogue.  We obtain approximately the same 2PCF results with our catalog and mask,  as in \cite{Siewert:2019} with 1 mJy flux cut and their mask-d. The theoretical curve at angular scales less than 0.1 degree is slightly lower than the measured values, presumably due to the multiple-component nature of some radio sources in the catalog \citep{Blake:2004}. We have followed \cite{Siewert:2019} and applied the same python package TreeCorr\footnote{\href{http://github.com/rmjarvis/TreeCorr/}{http://github.com/rmjarvis/TreeCorr/}} (Version 4.0) \citep{Jarvis:2004} and the same parameter settings for 2PCF estimation. Following \cite{Siewert:2019} we next divide the LoTSS DR1 catalogue into three patches `left', `center' and `right'. The  RA ranges for these patches are  (161, 184), (184, 208), and (208, 230) degrees, respectively. The calculated angular correlation functions are shown in figure \ref{fig:2pcf}. We find that the center area fits the theory best. The slight mismatch between `left', `center' and `right' patch results is an indication of large scale density fluctuation systematics present in the data. This should be considered as a motivation for refining the LoTSS DR1 pipeline \citep{Shimwell:2019}. 

\begin{figure}
\centering
\includegraphics[width=0.5\textwidth]{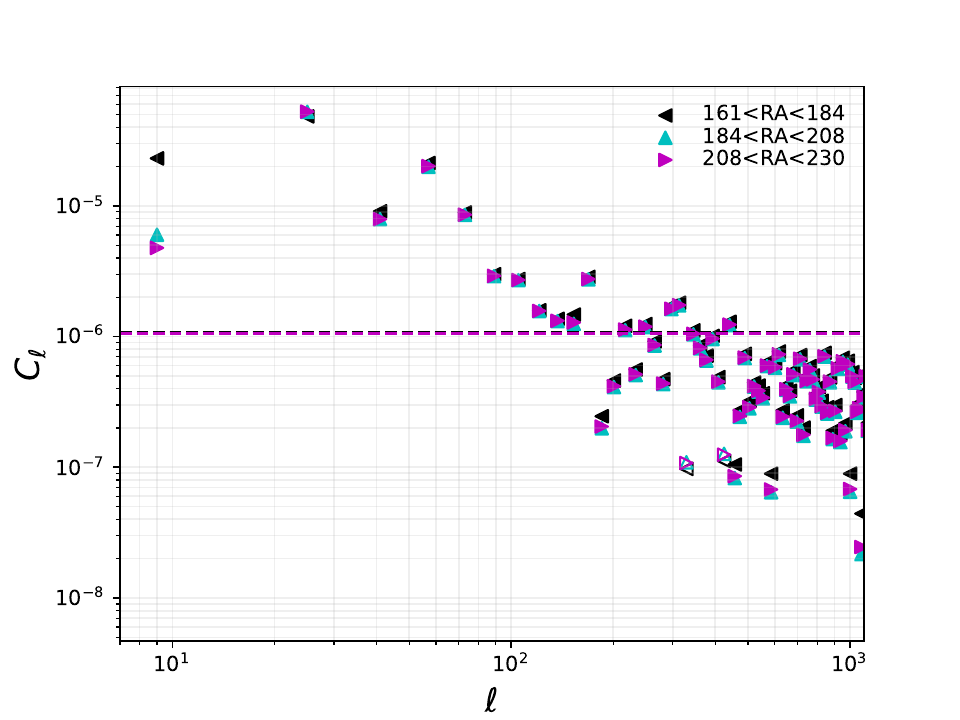}
\caption{Angular power spectrum recovery of LoTSS DR1 catalogues in different patches of the sky.  The recovered power spectrum from different right ascension patches is approximately the same above $\ell>10$. The dashed lines show the shot-noise. }
\label{fig:clRAcut}
\end{figure}

Large-scale systematics in real space correspond to low multipoles of angular power spectrum $C_{\ell}$. So to examine the effect more closely we calculate the angular power spectrum for these three right ascension patches. We show the results in figure \ref{fig:clRAcut} and note that $C_{\ell}$ recovery from different right ascension patches are almost the same above $\ell>10$. We recognize that $\ell =10$ corresponds to $\approx 20^\circ$ scale, which is approximately the right ascension patch width considered. We note that we only consider data above  $\ell=100$, and for this $\ell$ range we do not see any large scale right ascension dependence.

\section{Galaxy bias and $N(z)$ estimate}
\label{sc:bias_nz}
\begin{figure}
\includegraphics[width=0.5\textwidth]{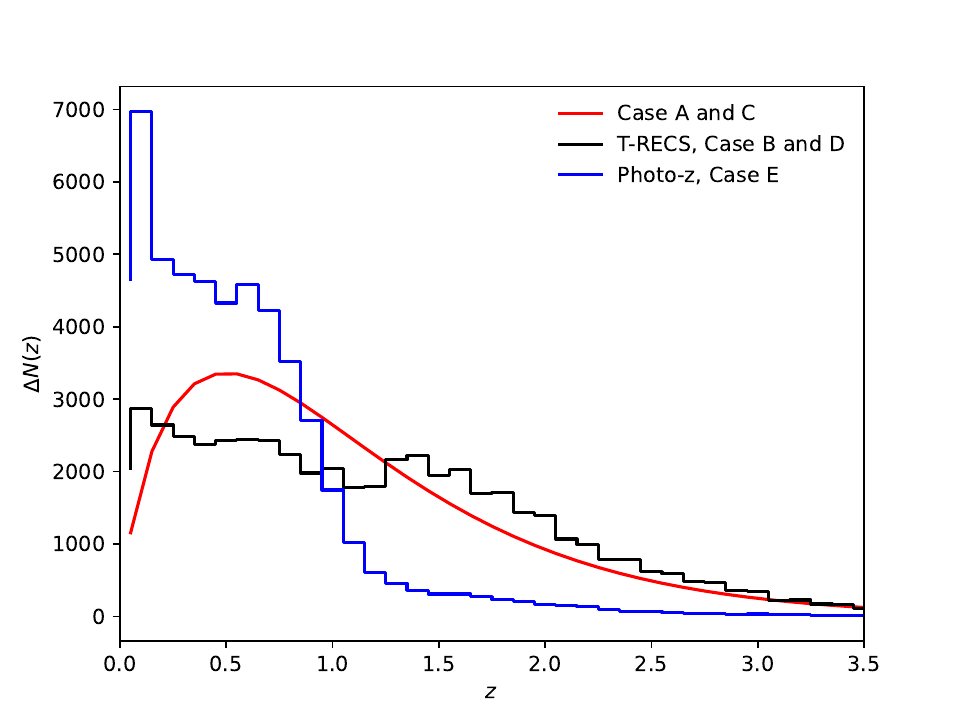}\\
\caption{Number of radio sources per redshift bin $\Delta z=0.1$ of available z for flux density thresholds $S>1$ mJy. The redshift distributions considered for fitting $\Lambda$CDM cosmology are also shown.}
\label{fig:photoz}
\end{figure}

The bias $b(z)$ for each galaxy population is different and depends on the dark matter halo mass hosting the galaxy type \citep{Mo:1996}. The angular power spectrum of LoTSS DR1 in figure \ref{fig:cl_recovery} is apparently  a close match with $\Lambda$CDM estimates considering NVSS galaxies bias and redshift distribution, $N(z)$. However, in reality, the bias and $N(z)$ for LoTSS DR1 galaxies could be significantly different in comparison with values obtained for NVSS galaxies with flux above $10$ mJy (NVSS completeness) at $1.4$ GHz \citep{Adi:2015nb,Tiwari:2016adi}. The NVSS is dominantly radio loud AGNs, Fanaroff–Riley type I (FR I) and FR II \citep{Fanaroff:1974}, whereas the LoTSS DR1 above  $1$ mJy contains star-forming galaxies along with radio loud AGNs,  FR I and FR II type \citep{Rivera:2017, Wilman:2008}. The LoTSS DR1 population is mixed and differs from NVSS, and thus the bias $b(z)$ and $N(z)$ could be significantly different. For $N(z)$, around half of the LoTSS DR1 sources have optical/near-IR  identification in Pan-STARRS/WISE and their photometric redshifts are available. The number count of these photo-identified galaxies in redshift bins ($\Delta z=0.1$) is shown in figure \ref{fig:photoz}. Note that the $N(z)$ in figure \ref{fig:photoz} only represents part of the LoTSS DR1 population; the true $N(z)$ of the full LoTSS DR1 catalog could be significantly different. Nevertheless, the available photo-redshifts can be assumed to loosely represent the LoTSS DR1 galaxies' redshift distribution. Furthermore, there are a few more reasonable $N(z)$ templates we may assume for LoTSS DR1, namely the $N(z)$ obtained for NVSS in \cite{Adi:2015nb} and the $N(z)$ presented in  the Tiered Radio Extragalactic Continuum Simulation (T-RECS, \citealt{Bonaldi:2018}).%%
Assuming a $\Lambda$CDM cosmology we seek to constrain the galaxy bias $b(z)$ and radial distribution $N(z)$ of LoTSS DR1 galaxies by fitting the measured angular power spectrum in figure \ref{fig:dataCl} and the observed $N(z)$ photo-$z$ histogram in figure \ref{fig:photoz}. For a given bias, $b(z)$ and $N(z)$ we calculate the theoretical (model) angular power spectrum $\clth$. Next, we obtain the likelihood, i.e. the probability to have observed data $\clcorr$ given the model, $\mathcal{P}( \clcorr|\bz,\nz) \propto \mathcal{L}(\clcorr) \propto \exp(-\frac{1}{2} (\vb{\clcorr}-\vb{\clth})^{T} \vb{\Sigma^{-1}}(\vb{\clcorr}-\vb{\clth}))$. Here $\vb{\Sigma}$ is the covariance matrix computed using mocks discussed in Section \ref{sc:covariance}. 
We make use of Bayes' probability theorem\footnote{$P(A|B) P(B)= P(B|A) P(A)$}, and write the model probability given the data, $\mathcal{P}( \bz,\nz|\clcorr)=\mathcal{P}( \clcorr|\bz,\nz) \times \mathcal{P}(\bz)$. Here $\mathcal{P}(\bz)$ is the prior probability of bias $\bz$.
\begin{figure}
    \centering
    \includegraphics[width=0.5\textwidth]{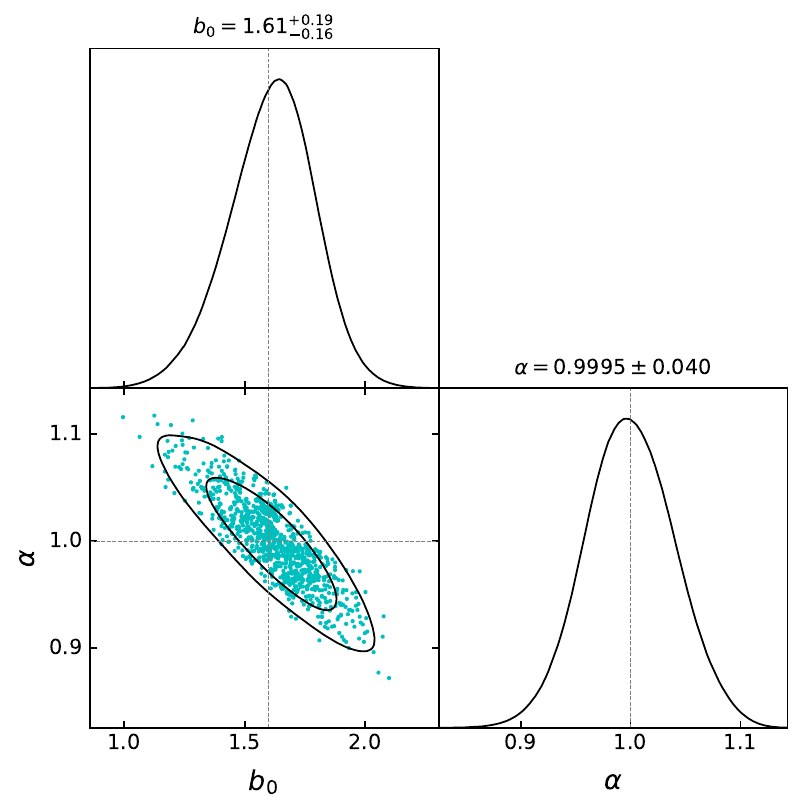}
    \caption{The pipeline recovers input bias parameters within one $\sigma$, dashed lines are input values. The figure in particular demonstrates the quadratic bias $b(z)=b_0+b_1 z+b_2 z^2$ recovery. For our fitting range, i.e. $\ell=200-1000$, we do not cover enough power spectrum shape to constrain $b_1$ and $b_2$ and the posterior probability  for these are almost the same as the prior information we assume.}
    \label{fig:mock_test}
\end{figure}
We assume different $N(z)$ templates and run an MCMC sampler to maximize model parameter probability, $\mathcal{P}(\bz,\nz|\clcorr)$, i.e. obtain the best fit to $\Lambda$CDM cosmology and deduce effective bias $b(z)$, given the $N(z)$ values. We fit one more model parameter, $\alpha$, to account for the multi-component offset to $\clcorr$ as discussed in sections \ref{sc:galaxyPk} and \ref{sc:cl_measure}. We conveniently use Cobaya \citep{Torrado:2020xyz} to perform Bayesian analysis and CosmoMC \citep{Lewis:2002ah,Lewis:2013hha} for MCMC sampling. 
With mock catalogs, we verify the above MCMC pipeline and recover the input bias within one sigma uncertainty for the $\ell=200$ to $1000$ range. The pipeline performance is demonstrated in figure \ref{fig:mock_test}. Above $\ell=1000$ the recovery from mocks is noisy and therefore we restrict our fits to $\ell \leq 1000$ for all cases discussed below. We would like to emphasize 
that the measured $C_\ell$ are projected quantities, and it is hard to extract tomographic information from them. The recovered bias is strongly dependent on the $N(z)$ assumption; the best we can do for now is to assume a variety of $N(z)$ models for the LoTSS catalog. We hope to have better $N(z)$ observations in future and thus better constraints on $b(z)$.

\begin{figure}
    \centering
    \includegraphics[width=0.5\textwidth]{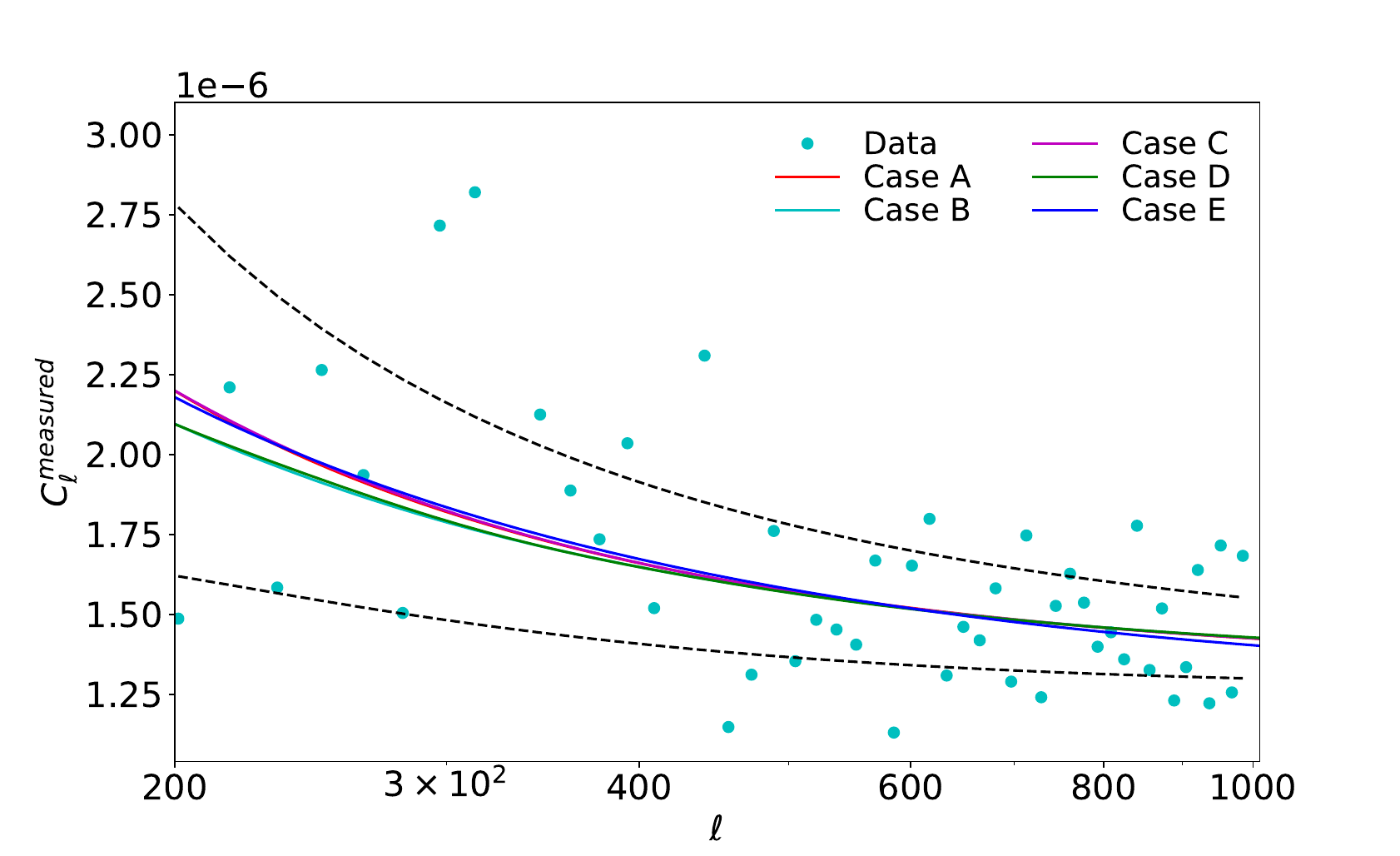}
    \caption{The best fit $\Lambda$CDM $C_\ell^{\rm measured}$ obtained using MCMC sampling with various sets of bias, $b(z)$ and $N(z)$ constraints and assumptions as discussed in Section \ref{sc:bias_nz}. The estimated observed pseudo-$C_\ell$ from data are also shown. The black dashed curves show the one sigma limit due to shot-noise and cosmic variance  following equation \ref{eq:dcl} for case A. For other cases the one sigma limit curves are approximately the same.} 
    \label{fig:best_Cl}
\end{figure}

{\bf Case A:}
Assume $N(z) \sim z^{0.74} \exp\left[-\left( \frac{z}{0.71}\right)^{1.1}\right]$ as given in \cite{Adi:2015nb}; we fit an effective redshift independent bias and shot-noise factor $\alpha$. We find that a bias  $b_{\rm eff}=1.90\pm{0.20}$ fits best with the data. The recovered best fit angular power spectrum is shown in figure \ref{fig:best_Cl}. The recovered shot-noise factor is $1.25\pm0.04$ which is consistent with $1.30$ as expected in section \ref{sc:cl_measure}.

\begin{figure}
    \centering
    \includegraphics[width=0.5\textwidth]{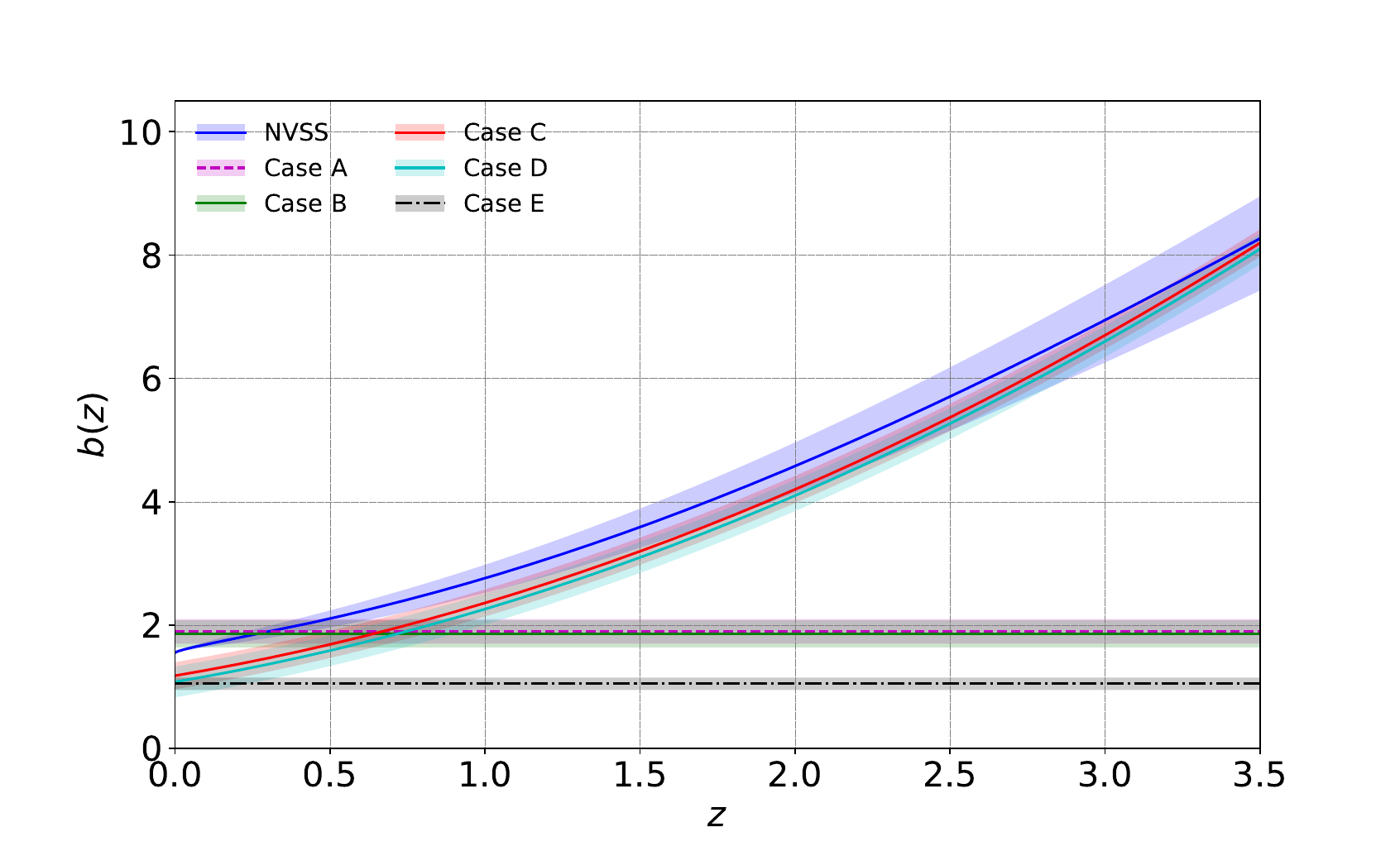}
    \caption{The bias, $b(z)$, recovered with MCMC Bayesian fitting for different cases. The best bias for NVSS radio AGNs obtained in \citep{Adi:2015nb} is also shown.}
    \label{fig:best_bz}
\end{figure}

{\bf Case B:}
We employ the  T-RECS \citep{Bonaldi:2018} simulation, and produce an $N(z)$ histogram considering the SFG+AGN population above 1 mJy flux at 150 MHz. The histogram is shown in figure \ref{fig:photoz}. We assume the T-RECS histogram\footnote{A cubic spline interpolation is used to make $N(z)$ smoother.} in figure \ref{fig:photoz} represents the LoTSS population and run the MCMC sampler to find the best redshift independent bias value. We obtain $b_{\rm eff}=1.86\pm{0.22}$ as best fit to the data. The best fit angular power spectrum and bias with one sigma uncertainty band  is shown in figures \ref{fig:best_Cl} and \ref{fig:best_bz}, respectively. 

\begin{figure}
    \centering
    \includegraphics[width=0.5\textwidth]{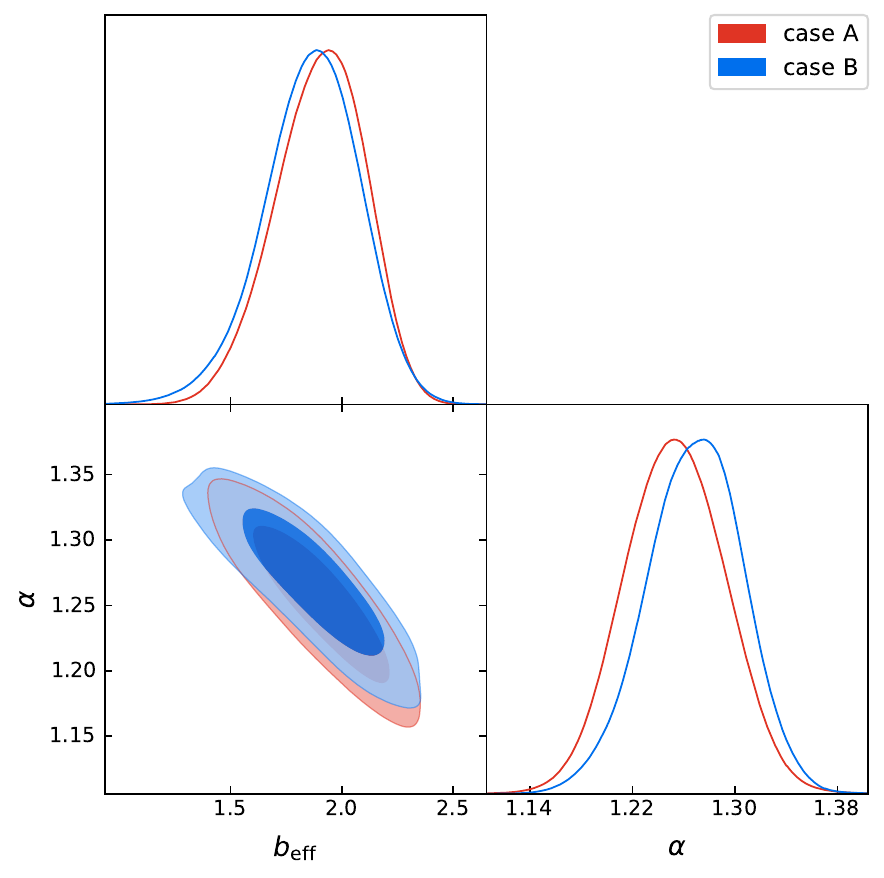}
    \caption{Case A and B: Posterior distributions of bias and shot-noise factor $\alpha$. $(\alpha -1)\times \sigma_0$ is the multi-component offset.}
    \label{fig:corner_caseAB}
\end{figure}

\begin{figure}
    \centering
    \includegraphics[width=0.5\textwidth]{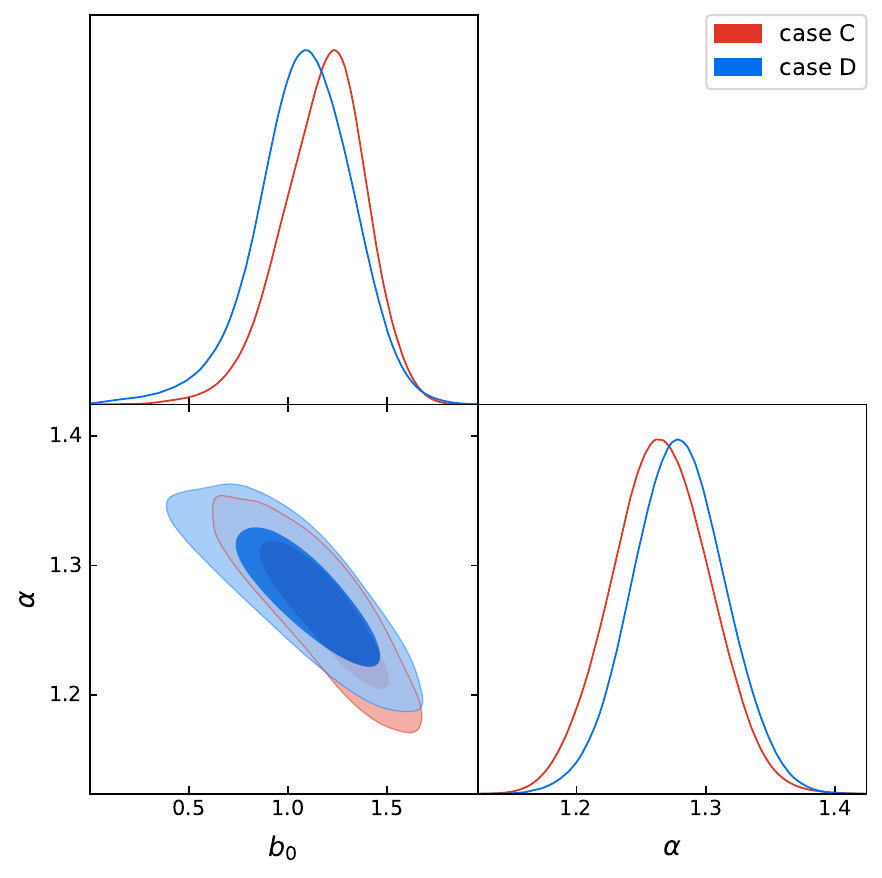}
    \caption{Case C and D: One and two-dimensional posterior distributions of quadratic bias free parameters $b_0$ and shot-noise factor $\alpha$. As the data is not sensitive to $b_1$ and $b_2$, we fix these to NVSS observed values. Other details same as in figure \ref{fig:corner_caseAB}. }
    \label{fig:corner_caseCD}
\end{figure}

{\bf Case C:} 
$N(z)$ same as in case A; we fit a quadratic bias $b(z)=b_0+b_1 z+b_2 z^2$. This is to test how well NVSS quadratic bias obtained in \cite{Adi:2015nb} fits to LoTSS DR1. 
We have noticed with mocks that the data is not sensitive to $b_1$ and $b_2$, we recover the same posterior probability  for these as the prior information we assume.
We fix these to NVSS observed values, i.e. $b_1=0.85$, $b_2=0.33$ and run the MCMC sampler over $b_0$ and noise+multi-component factor $\alpha$. The best-fit bias, $b(z)$, and angular power spectrum thus recovered are shown in figures \ref{fig:best_bz} and \ref{fig:best_Cl}, respectively. The MCMC sample and posterior distributions are shown in figure \ref{fig:corner_caseCD}. We do see a better fit with quadratic bias; the fit parameters AIC and BIC are slightly improved. All parameters, their prior information and fit parameters are given in table \ref{tb:results}.

{\bf Case D:}
We choose $N(z)$ to be the same as in case B, and fit a quadratic bias $b(z)=b_0+b_1z+b_2z^2$. 
Other details are the same as in case C. The recovered best bias, $b(z)$, and angular power spectrum are shown in figures \ref{fig:best_bz} and \ref{fig:best_Cl}, respectively. The MCMC sample and posterior distributions are shown in figure \ref{fig:corner_caseCD}.

{\bf Case E:}
Finally, we use observed photo-z redshifts which are available for onlyy about half of the radio sources. This $N(z)$ only  represents  part  of  the  LoTSS  DR1  population, but we include this case to check the sensitivity of the data to $N(z)$. We assume  a redshift independent bias, i.e. $b(z)=b_{\rm eff}$ and a shot-noise factor $\alpha$, and then fit with the recovered LoTSS DR1 angular power spectrum.  We obtain a best fit bias $b_{\rm eff}=1.05\pm{0.10}$.  The recovered best power spectrum is shown in figure \ref{fig:best_Cl}. 

\begin{figure}
    \centering
    \includegraphics[width=0.5\textwidth]{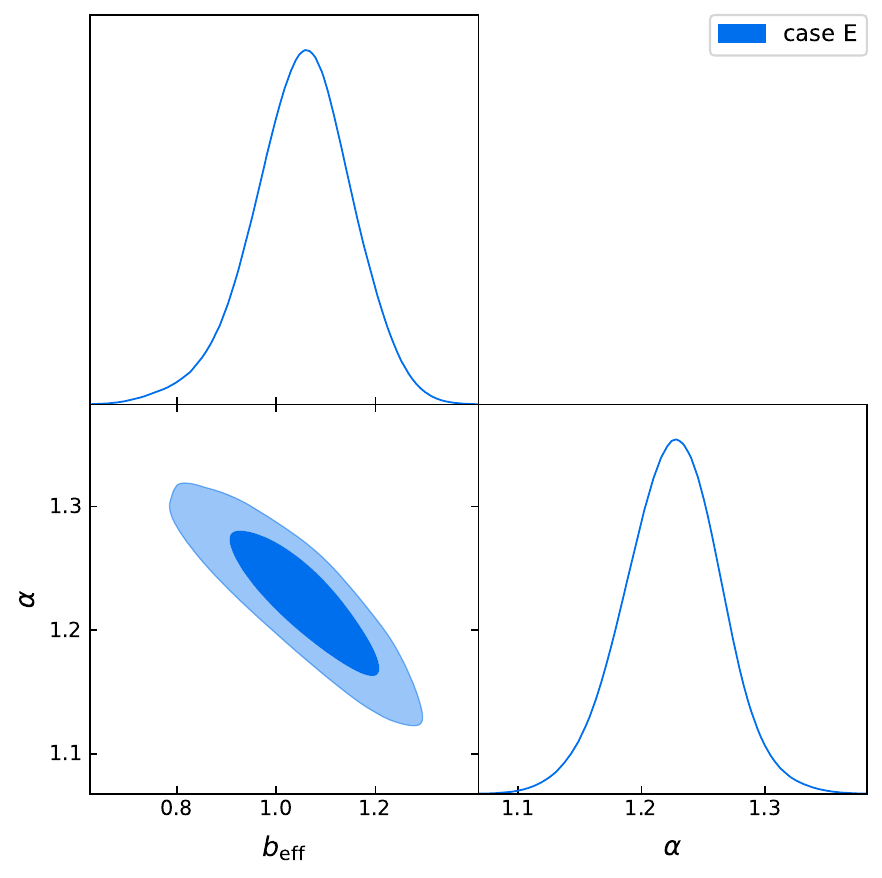}
    \caption{Case E: Posterior distributions of bias and shot-noise factor $\alpha$. Other details same as in figure \ref{fig:corner_caseAB}.}
    \label{fig:corner_caseE}
\end{figure}

\begin{deluxetable*}{c c c c c c c c}
\tablecaption{MCMC sampler results. The prior and maximum likelihood value of the model free parameters and uncertainties. The flat prior range is shown in square brackets.The reduced $\chi^2$, i.e. $\chi^2$ per degree of freedom, AIC and BIC are also listed for model fit comparison. \label{tb:results}}
\centering 
\tablehead{ & & Case A & Case B & Case C & Case D & Case E} 
\startdata
\multirow{3}*{$b_{_0}$} & prior     &    &    & $[0.0, 5.0]$ & $[0.0, 5.0]$ & \\
                        &&&&&&\\
                        & fit value & -- & -- & $1.18\pm{0.22}$ & $1.08\pm{0.25}$ & --\\
\vspace{0.1cm}\\

$b_{_1}$ & fix     &    &  & $0.85$ & $0.85$ & \\
\vspace{0.1cm}\\

$b_{_2}$ & fix     &    &    & $0.33$ & $0.33$ &  \\
\vspace{0.1cm}\\

\multirow{3}*{$b_{\rm eff}$} & prior     & [0.5, 5.0]   & [0.1, 5.0] & & & [0.1, 5.0]\\
                        &&&&&&\\
                        & fit value & $1.90\pm{0.20}$ & $1.86\pm{0.22}$ & -- & -- & $1.05\pm{0.10}$\\
\vspace{0.1cm}\\
\multirow{3}*{$\alpha$} & prior     & [0.01, 3.0]   & [0.01, 3.0] & [0.01, 3.0] & [0.01, 3.0] & [0.01 3.0]\\
                        &&&&&&\\
                        & fit value & $1.25\pm{0.04}$ & $1.27\pm{0.04}$ & $1.26\pm{0.04}$ & $1.28\pm{0.04}$ & $1.22\pm{0.04}$\\
\vspace{0.1cm}\\

$\bm{\chi^2}$/\bf{dof}& & \bf{1.18}    & \bf{1.36} & \bf{1.18} & \bf{1.36} & \bf{1.39}\\
\vspace{0.1cm}\\

\bf{AIC}&    & \bf{60.65} & \bf{69.28} & \bf{60.49} & \bf{69.06} &\bf{70.66} \\
\vspace{0.1cm}\\

\bf{BIC}&    & \bf{64.47} & \bf{73.11} & \bf{64.32} & \bf{72.88} &\bf{74.48}\\
\end{deluxetable*}

\section{Summary}
\label{sc:summary}

We have explored the first data release of the LoTSS survey for cosmological studies. We have generated detailed data mocks and constructed the covariance matrix for angular power spectrum recovery. With our mocks, 
we were able to customise the settings of the pseudo angular power recovery algorithm for LoTSS. The LoTSS DR1 contains dominantly star-forming galaxies and a significant fraction of AGNs; therefore the resultant bias and radio distribution of galaxies can't  be naively assumed to be the same as seen with other radio surveys with high flux limit, e.g. the NVSS, where the catalog almost entirely  consists of AGNs. We have explored the galaxy bias and possible radial distribution profiles for LoTSS galaxies and have presented their fits to data. We find that the data above integrated flux density 1 mJy and with mask-z has a reasonable angular power spectrum. Furthermore, we consider the multi-component fraction of radio galaxies as an independent fit parameter and recover its value. For convenience, the summary of different cases, the recovered bias  with fit quality parameters, i.e. $\chi^2/$dof, and model selection criteria parameters, the Akaike's information criterion (AIC) and Bayesian information criterion (BIC), are listed in Table \ref{tb:results}. 

We have tried fitting  the best bias and $N(z)$ profiles from NVSS \citep{Adi:2015nb,Tiwari:2016adi} to LoTSS and obtain that the bias for LoTSS DR1 is slightly lower in comparison with NVSS if we assume NVSS $N(z)$ (Case C).
Our bias values {for cases A, B, C and D} are {similar to}  \cite{Alonso:2020jcy}, obtained using CMB lensing and LoTSS DR1 data. Considering redshift information dominantly determined by available photo-$z$ information (Case E), we obtain a lower bias value close to 1; this low value of bias presumably demonstrates that i) the data is sensitive to the $N(z)$ distribution and ii) the photo-$z$ distribution for the limited number of sources does not represent a reasonable $N(z)$ for the full LoTSS DR1 population.
Considering different $N(z)$ templates and bias $b(z)$ forms, with an MCMC sampler, we  find that the best fit result (best reduced-$\chi^2$, AIC, and BIC) is with Case C i.e. by assuming $N(z)$ given in \cite{Adi:2015nb} and a redshift dependent quadratic bias. The reduced $\chi^2$ of all models is slightly higher than 1. This may indicate some residual systematics.  We expect LoTSS DR2 to have better control of systematics and thus to provide better constraints on cosmology and galaxy bias. The multi-component fraction for LoTSS DR1 is not very settled and we consider this as an independent parameter. We find that a constant offset to $C_\ell$ equal to $\approx 0.25$ times shot-noise is needed. The performance of our MCMC sampler has been verified by recovering model input bias, $b(z)$,  by fitting pseudo-$C_\ell$ mean from mocks.  

To compare our results  with \cite{Siewert:2019}, we have also presented measurements for the 2PCF for our catalog. We obtain approximately the same results as in \cite{Siewert:2019}. The 2PCFs calculated for broad right ascension patches differ; this potentially indicates large-scale systematics in the data. These large-scale anomalies in the real space clustering signal correspond to low multipoles, approximately $\ell<10$. For our fits we have considered the power spectrum above multipole $\ell=200$ and for this $\ell$ range we have verified that the recovered power spectrum is approximately the same for all three right ascension patches.  

\section{Conclusion and Discussion} 
\label{sc:discussion}

We have presented detailed mocks and projected clustering statistics of point sources from LoTSS first data release, based on the observation of 424 square degrees of the sky at $\sim 150$MHz. We assume different possible redshift templates, and rum MCMC sampler to obtain best galaxy bias and multi-component fraction  for LoTSS DR1 above 1 mJy. With 2-point correlation statistics, we notice somewhat different clustering in large right-ascension patches, presumably indicating large-scale flux calibration systematics in data. The angular power spectrum $C_\ell$, the spherical harmonic decomposition of the fluctuations in angular space, efficiently separates the scale-dependent real space number density fluctuations. The catalog contains multiple entries for a significant number of sources, resulting in a constant offset to the measured angular power spectrum. With the multi-component correction,  we see a good match of LoTSS DR1 measured $C_\ell$ with standard $\Lambda$CDM theory. The data above 1 mJy and with mask-z results in consistent $C_\ell$s and arguably qualifies for cosmological analysis.

The LOFAR observations are at relatively low radio frequency and the strength of the systematic effects affecting the data is significantly higher than high frequency surveys, e.g., VLA NVSS, and thus the calibration and source cataloging is challenging. Even so, the data calibration pipeline by \cite{Shimwell:2019} delivers a workable source catalog for cosmology. The LOFAR team will continue to improve on flux calibration and cataloging. Furthermore, we need better estimates for redshifts; the present partial availability of photometric redshifts is not reliably representing the LoTSS population. Ideally, above the LoTSS completeness limit we need accurate redshifts for all galaxies over a small sky patch to obtain a complete redshift distribution of LoTSS galaxies \citep{Duncan:2020}. With improved calibration, large sky coverage, and higher number density,  the upcoming catalogs from LOFAR are expected to be a significant improvement over current low frequency radio catalogs, e.g., the TGSS from GMRT and GLEAM from MWA. The upcoming LoTSS catalogs are an excellent opportunity for cosmological studies in the low radio frequency regime.   

\section{Acknowledgments}
 We thank Thilo Siewert for helping with the LoTSS DR1 catalog and masks. PT acknowledges the support of the RFIS grant (No. 12150410322) by the National Natural Science Foundation of China (NSFC).  PT, RZ, JZ and GBZ are supported by the National Key Basic Research and Development Program of China (No. 2018YFA0404503) and NSFC Grants 11925303 and 11720101004, and a grant of CAS Interdisciplinary Innovation Team. JZ is supported by Chinese Scholarship Council (CSC) and Science and Technology Facilities Council (STFC) for visiting University of Portsmouth. Some of the results in this paper have been derived using the HEALPix \citep{Gorski:2005} package.

 LOFAR data products were provided by the LOFAR Surveys Key Science project (LSKSP; \href{https://lofar-surveys.org/}{https://lofar-surveys.org/}) and were derived from observations with the International LOFAR Telescope (ILT). LOFAR \citep{Haarlem:2013} is the Low Frequency Array designed and constructed by ASTRON. It has observing, data processing, and data storage facilities in several countries, which are owned by various parties (each with their own funding sources), and which are collectively operated by the ILT foundation under a joint scientific policy. The efforts of the LSKSP have benefited from funding from the European Research Council, NOVA, NWO, CNRS-INSU, the SURF Co-operative, the UK Science and Technology Funding Council and the J\"{u}lich Supercomputing Centre.
\bibliographystyle{apj}
\bibliography{main}
\end{document}